\documentclass[11pt, a4paper]{article}
\usepackage{fullpage}

\usepackage{belov_paper}
\usepackage{amssymb}
\usepackage{graphicx}
\usepackage{calc}

\usepackage{euscript}

\title{Quantum Algorithms for Classical Probability Distributions}
\author{
  Aleksandrs Belovs\thanks{Faculty of Computing, University of Latvia.}
}
\date{}

\newcommand{\qqhence}{\qquad\text{hence}\qquad}

\makeatletter
\newcommand*{\bigboxplus}{%
  \DOTSB
  \mathop{\vphantom{\bigoplus}\mathpalette\matt@bigboxplus\relax}%
  \slimits@
}
\newcommand\matt@bigboxplus[2]{%
  \vcenter{\m@th\hbox{\resizebox{\widthof{$#1\bigoplus$}}{!}{$\boxplus$}}}%
}
\makeatother

\def\reg#1{\mathsf{#1}}

\begin{document}
\maketitle

\begin{abstract}
We study quantum algorithms working on classical probability distributions.  We formulate four different models for accessing a classical probability distribution on a quantum computer, which are derived from previous work on the topic, and study their mutual relationships.

Additionally, we prove that quantum query complexity of distinguishing two probability distributions is given by their inverse Hellinger distance, which gives a quadratic improvement over classical query complexity for any pair of distributions.

The results are obtained by using the adversary method for state-generating input oracles and for distinguishing probability distributions on input strings.
\end{abstract}

\mycommand{dH}{\mathop{\mathrm{d_H}}}

\section{Introduction}
It is customary for a quantum algorithm to receive its input and produce its output in the form of a classical string of symbols, quantized in the form of an oracle.
This is purely classical way to store information, and, given intrinsic quantum nature of quantum algorithms, this might be not the best interface for many tasks.
Moreover, even \emph{classical} algorithms make use of other interfaces as well.
For instance, classical algorithms can receive and produce \emph{samples} from some probability distribution.
In this paper we study quantum algorithms working with classical probability distributions.

\paragraph{Models.}
We analyse previously used models of accessing classical probability distributions by quantum algorithms.  We prove and conjecture some relations between them.
We give more detail in \rf{sec:models}, but for now let us very briefly introduce the models.

In one of the models, used in, e.g., \cite{bravyi:testingDistributions, chakraborty:testingDistributions, montanaro:frequencyMoments, li:entropyEstimation}, the probability distribution is encoded as a frequency of a symbol in a given input string, which the quantum algorithm accesses via the standard input oracle.
In another model, e.g., \cite{bshouty:learningDNF, aharonov:adiabaticStateGeneration, atici:improvedBoundsLearning}, the input probability distribution is given through a quantum oracle that prepares a state in the form $\sum_{a} \sqrt{p_a}\ket|a>$.
Finally, one more model, used in~\cite{montanaro:MonteCarlo, hamoudi:chebyshev, gilyen:distributionalTesting}, is similar but with additional state tensored with each $\ket|a>$.

This is the latter model that we champion in this paper.
We find this model particularly relevant because of our believe that an input oracle should be easily interchangeable with a quantum subroutine, see discussion in~\cite{montanaro:MonteCarlo}.
It is relatively easy to see what it means for a quantum algorithm to output a probability distribution: just measure one of the registers of its final state.
The latter model precisely encompasses all such subroutines.
We conjecture that this model is equivalent to the first model, see also~\cite{gilyen:distributionalTesting}, where a similar conjecture is made.

%Additionally, our first mentioned model is a special case of the latter model.
%We conjecture that these two models are equivalent in the sense replacing an oracle of one kind with an oracle of another kind can only change query complexity of the problem by at most a constant factor.  
%Our second mentioned model, on the other hand, is too strong and can result in much faster algorithms.

\paragraph{Distinguishing two Probability Distributions.}
Additionally, we study the problem of distinguishing two probability distributions.  
This might be the most fundamental problem one can formulate in these settings.
Given two fixed probability distributions $p$ and $q$, and given an input oracle encoding one of them, the task is to detect which one, $p$ or $q$, the oracle encodes.
To the best of our knowledge, this particular problem has not been studied in quantum settings, although similar problems of testing the distance between two distributions~\cite{bravyi:testingDistributions} and testing whether the input distribution is equal to some fixed distribution~\cite{chakraborty:testingDistributions} have been already studied.

Classically one needs $\Theta\sA[1/\dH(p,q)^2]$ samples to solve this problem for any $p$ and $q$, where $\dH$ stands for Hellinger distance.  
This result is considered ``folklore'', see, e.g.~\cite[Chapter 4]{BarYossef:phd}.
We prove that for any $p$ and $q$ and for any of the models of access described above, query complexity of this problem is $\Theta(1/\dH(p,q))$.
This constitutes quadratic improvement over classical algorithm for \emph{any} pair of distributions $p$ and $q$.
Moreover, our algorithm also admits a simple low-level implementation, which is efficient assuming the distributions $p$ and $q$ can be efficiently processed.

\paragraph{Techniques.}
Our main technical tool for proving the upper bound is the version of the adversary bound for state-generating oracles, which is a special case of the adversary bound for general input oracles~\cite{belovs:variations}.
It is stated in the form of a relative $\gamma_2$-norm and generalises the dual formulation of the general adversary bound~\cite{reichardt:spanPrograms, reichardt:advTight} for function evaluation, as well as for other problems~\cite{ambainis:symmetryAssisted, lee:stateConversion}.
The dual adversary bound has been used rather successfully in construction of quantum algorithms, as in terms of span programs and learning graphs~\cite{belovs:learning, lee:learningSubgraphs, belovs:learningClaws, belovs:learningKDist, jarret:connectivity}, as in an unrelated fashion~\cite{belovs:learningSymmetricJuntas, belovs:gappedGroupTesting}.
Our work gives yet another application of these techniques for construction of quantum algorithms.  

Our upper bound naturally follows from the analysis of the $\gamma_2$-norm optimisation problem associated with the task.
We also compare our techniques with more standard ones involving quantum rejection sampling and amplitude amplification in the spirit of~\cite{hamoudi:chebyshev} and show that our techniques give a slightly better result.

As for the lower bound, we make use of the version of the adversary bound from~\cite{belovs:merkle}.
This is a simple generalisation of the primal version of the general adversary bound~\cite{hoyer:advNegative} for function evaluation, and it is tailored for the task we are interested in: distinguishing two probability distributions on input strings.
Our lower bound is surprisingly simple and gives a very intuitive justification of the significance of Hellinger distance for this problem.

\section{Preliminaries}
We mostly use standard linear-algebraic notation.  
We use ket-notation for vectors representing quantum states, but generally avoid it.
We use $A^*$ to denote conjugate operators (transposed and complex-conjugated matrices).
For $P$ a predicate, we use $1_P$ to denote 1 if $P$ is true, and 0 if $P$ is false.
We use $[n]$ to denote the set $\{1,2,\dots,n\}$.

It is unfortunate that the same piece of notation, $\oplus$, is used both for direct sum of matrices and direct sum of vectors, which is in conflict with each other if a vector, as it often does, gets interpreted as a column-matrix.
Since we will extensively use both these operations in this paper, let us agree that $\boxplus$ denotes direct sum of vectors, and $\oplus$ always denotes direct sum of matrices.
Thus, in, particular, for $u,v\in\bR^m$, we have
\[
u\boxplus v =
\begin{pmatrix}
u_1\\\vdots\\u_m\\v_1\\\vdots\\v_m
\end{pmatrix}
\qqand
u\oplus v =
\begin{pmatrix}
u_1&0\\\vdots&\vdots\\u_m&0\\0&v_1\\\vdots&\vdots\\0&v_m
\end{pmatrix}
.
\]
We often treat scalars as $1\times1$-matrices which may be also thought as vectors.

\subsection{Relative \texorpdfstring{$\gamma_2$}{gamma2}-norm}
\label{sec:gamma2}
In this section, we state the relative $\gamma_2$-norm and formulate some of its basic properties.  All the results are from~\cite{belovs:variations}.

\begin{defn}[Relative $\gamma_2$-norm]
\label{defn:matrixGamma2}
Let $\cX_1$, $\cX_2$, $\cZ_1$ and $\cZ_2$ be vector spaces, and $D_1$ and $D_2$ be some sets of labels.
Let $A = \{A_{xy}\}$ and $\Delta = \{\Delta_{xy}\}$, where $x\in D_1$ and $y\in D_2$, be two families of linear operators: $A_{xy}\colon \cZ_2\to\cZ_1$ and $\Delta_{xy}\colon \cX_2\to\cX_1$.
The \emph{relative $\gamma_2$-norm}, 
\[
\gamma_2(A | \Delta) = \gamma_2(A_{xy} \mid \Delta_{xy})_{x\in D_1,\; y\in D_2},
\]
is defined as the optimal value of the following optimisation problem, where $\Upsilon_x$ and $\Phi_y$ are linear operators,
\begin{subequations}
\label{eqn:matrixGamma2}
\begin{alignat}{2}
&\mbox{\rm minimise} &\quad& \max \sfigB{ \max\nolimits_{x\in D_1} \norm|\Upsilon_x|^2, \max\nolimits_{y\in D_2} \norm|\Phi_y|^2 } \\
& \mbox{\rm subject to}&&  
A_{xy} = \Upsilon_x^* (\Delta_{xy}\otimes I_{\cW}) \Phi_y \quad \text{\rm for all $x\in D_1$ and $y\in D_2$;}  \label{eqn:matrixGamma2Condition}\\
&&& \text{$\cW$ is a vector space,\quad 
$\Upsilon_x\colon \cZ_1\to \cX_1\otimes\cW$,\quad $\Phi_y\colon \cZ_2\to \cX_2\otimes\cW$.}
\end{alignat}
\end{subequations}
\end{defn}

This is a generalisation of the usual $\gamma_2$-norm, also known as Schur (Hadamard) product operator norm~\cite{bhatia:positive}.
%Also, the usual dual general adversary bound for evaluation a function $f\colon D\to R$ with $D\subseteq [\ell]^n$ admits the following expression as a relative $\gamma_2$-norm: $\gamma_2\sA[1_{f(x)\ne f(y)} \midA \bigoplus_j 1_{x_j\ne y_j}]_{x,y\in D}$, where $\bigoplus$ stands for direct sum of matrices.

In a quantum algorithm with general input oracles, the input oracle performs some unitary operation $O_x$ on some fixed Hilbert space, where $x$ ranges over some set $D$ of labels, and the algorithm has to perform a unitary $V_x$ on some specified part of its work-space.
The algorithm knows in advance all possible $O_x$ and which $V_x$ corresponds to each $O_x$, but it does not know which $O_x$ it is given in a specific execution.
The adversary bound corresponding to this problem is $\gamma_2\sA[V_x-V_y\mid O_x-O_y]_{x,y\in D}$.  
This bound is \emph{semi-tight}: it is a lower bound on the exact version of the problem and an upper bound on the approximate version.

The $\gamma_2$-norm formalism is modular in the sense that the general task of implementing a unitary can be replaced by something more specific.  For instance, assume that our task is to evaluate a function $f(x)$.  
Then the adversary bound reads as $\gamma_2\sA[1_{f(x)\ne f(y)}\mid O_x-O_y]_{x,y\in D}$.  In this case, the bound is tight: it is also a lower bound on the approximate version of the problem.

As another example, consider the standard input oracle $O_x$ encoding a string $x\in[q]^n$.
It works as $O_x\colon \ket|i>\ket|0>\mapsto \ket|i>\ket|x_i>$, which 
can be seen as a direct sum of oracles performing transformation $\ket|0>\mapsto\ket|x_i>$.
Using the modular approach, the corresponding adversary bound becomes $\gamma_2\sA[1_{f(x)\ne f(y)} \mid \bigoplus_j 1_{x_j\ne y_j}]_{x,y\in D}$, where $\bigoplus$ stands for direct sum of matrices (resulting in a diagonal matrix).  This is equivalent to the usual version of dual adversary for function evaluation (up to a constant factor).

Now we consider state-generating input oracles\footnote{
        The results below will appear in an updated version of~\cite{belovs:variations} (to appear).  For completeness, we place the proof of \rf{thm:gamma2StatePreparing} in \rf{app:proofgamma2}, as it should appear in the new revision of~\cite{belovs:variations}.
}.
In this case, the input to the algorithm is given by a state $\psi\in\bC^m$, and the algorithm should work equally well for any unitary performing the transformation $O\colon \ket|0> \mapsto \ket|\psi>$.
Without loss of generality, we may assume that $e_0 = \ket|0>$ is orthogonal to $\bC^m$, thus the operator $O$ above works in $\bC^{m+1}$.

The corresponding $\gamma_2$-object can be defined in two alternative ways:
\[
L_\psi = \psi e_0^* + e_0\psi^*
\qquad\text{or}\qquad
L_\psi = \psi \oplus \psi^*.
\]
In the second expression, $\psi$ is an $m\times 1$-matrix and $\psi^*$ is a $1\times m$-matrix, the resulting matrix being of size $(m+1)\times(m+1)$.
In the case of a function-evaluation problem, the corresponding adversary bound is $\gamma_2\sA[ 1_{f(x)\ne f(y)} \mid L_{\psi_x} - L_{\psi_y} ]_{x,y\in D}$.

Let us also state the version of the adversary bound for the decision problem with state-generating input oracles.
This is the version we will use further in the paper.
Assume we have a collection of states $\psi_x\in \cX$ for $x\in D_0$, and a collection of states $\psi_y\in \cX$ for $y\in D_1$.
The task is to distinguish the two classes of states.  Let $D=D_0\cup D_1$.
Using the general case, we obtain the following version of the adversary bound.

\begin{thm}
\label{thm:gamma2StatePreparing}
The quantum query complexity of the decision problem with state-generating oracles as above is equal to $\gamma_2\sA[1 \mid  L_{\psi_x}-L_{\psi_y} ]_{x\in D_0, y\in D_1}$ up to a constant factor.
\end{thm}

An explicit optimisation problem for $\gamma_2\sA[1 \mid  L_{\psi_x}-L_{\psi_y} ]_{x\in D_0, y\in D_1}$ is given by
\begin{equation}
\label{eqn:gamma2StatePreparing}
\begin{aligned}
\text{minimise}&\quad&& \max\nolimits_{z\in D} \sA[\|u_z\|^2 + \|v_z\|^2]\\
\text{subject to}&&& \ipA<v_x, (\psi_x - \psi_y)\otimes u_y> + \ipA<(\psi_x-\psi_y)\otimes u_x, v_y>=1 &&\forall x\in D_0, y\in D_1;\\
&&& u_z \in \cW,\quad v_z\in \cX\otimes \cW
&&\forall z\in D.
\end{aligned}
\end{equation}

This result follows from general results of~\cite{belovs:variations}, see \rf{app:proofgamma2}. We also give a stand-alone implementation and analysis of the corresponding quantum algorithm in \rf{app:gamma2StatePreparing}.

\section{Models}
\label{sec:models}
In this section we formally define four different models how a quantum algorithm can access a classical probability distribution $p=(p_a)_{a\in A}$.  These models were briefly explained in the introduction.
We would like to understand relations between them, and, ideally, prove some equivalences between them.

\begin{itemize}
\item[(i)] 
A standard input oracle encoding a string $x\in A^n$ for some relatively large $n$, where $p_a$ is given as the frequency of $a$ in $x$:
\[
p_a = \frac1n \absB|\{ i \mid x_i = a \}|.
\]

\item[(ii)] 
A standard input oracle encoding a string $x\in A^n$ for some relatively large $n$, where each $x_i$ is drawn independently at random from $p$.

\item[(iii)] A quantum procedure that generates the state
\begin{equation}
\label{eqn:OracleII}
\mu_p = \sum_a \sqrt{p_a} \ket|a> = \bigboxplus_a \sqrt{p_a}.
\end{equation}

\item[(iv)] A quantum procedure that generates a state of the form
\begin{equation}
\label{eqn:OracleI}
\sum_a \sqrt{p_a} \ket|a>\ket|\psi_a> = \bigboxplus_a \sqrt{p_a}\psi_a,
\end{equation}
where $\psi_a$ are arbitrary unit vectors.

\end{itemize}

As mentioned in the introduction, model (i) is used in~\cite{bravyi:testingDistributions, chakraborty:testingDistributions, montanaro:frequencyMoments, li:entropyEstimation}.
It has a downside that the probabilities $p_a$ must be multiples of $1/n$.  All other models are free from this assumption.

Model (ii) seems like the most obvious way to encode probability distribution as a classical string, which a quantum algorithm can later gain access to.  Up to our knowledge, this model has not been previously used.  
It has a downside that the distribution $p$ is encoded as a probability distribution over possible input strings, which is not usual for quantum algorithms.  The acceptance probability of the quantum algorithm depends both on the randomness introduced by the algorithm and the randomness in the input.

Model (iii) is the one used in~\cite{bshouty:learningDNF, aharonov:adiabaticStateGeneration, atici:improvedBoundsLearning}.  And model (iv) is used in~\cite{montanaro:MonteCarlo, hamoudi:chebyshev, gilyen:distributionalTesting}.  Both of these two models assume that the input oracle prepares a quantum state, which again is not very common for quantum algorithms.

\begin{prp}
\label{prp:relations}
We have the following relations between these models.
\begin{itemize}
\item[(a)] Models (i) and (ii) are equivalent assuming $n$ is large enough.  
More precisely, no quantum algorithm can distinguish models (i) and (ii) encoding the same probability distribution unless it makes $\Omega(n^{1/3})$ queries.
\item[(b)] Model (iv) is more general than model (i).
\item[(c)] Model (iv) is strictly more general than model (iii).  
This means there exist problems where model (iii) allows substantially smaller query complexity than model (iv).
\end{itemize}
\end{prp}

\pfstart
We leave (a) for the end of the proof, and let us start with (b).  Note that using one query to the input oracle of model (i), it is possible to prepare that state
\[
\frac{1}{\sqrt n}\sum_i \ket|i>\ket|x_i>
= \sum_{a\in A} \skC[\frac{1}{\sqrt n}\sum_{i: x_i=a} \ket|i>]\otimes\ket|a>,
\]
which is a legitimate input state in model (iv) if one swaps the registers.
\medskip

Now let us prove (c).
It is obvious that model (iv) is more general than model (iii).
To prove that (iii) cannot simulate (iv), consider the collision problem~\cite{brassard:collision}.
In this problem, a function $f\colon [n]\to[n]$ is given, and one has to distinguish whether $f$ is 1-to-1 or 2-to-1.
In terms of model (i), this boils down to distinguishing a probability distribution $p$ which is uniform on $[n]$ from a probability distribution $q$ which is uniform on half of $[n]$.

In model (iii), this problem can be solved in $O(1)$ queries because the state $\mu_p$ as in~\rf{eqn:OracleII} has inner product $1/\sqrt{2}$ with all $\mu_q$.
On the other hand, by~\cite{shi:collisionLower, ambainis:collisionLower}, quantum query complexity of this problem in model (i) is $\Omega(n^{1/3})$.
As model (iv) is more general than model (i), this gives the required lower bound.
\medskip

To prove (a), we show that if one can distinguish models (i) and (ii), one can distinguish a random function from a random permutation, and the result follows from the lower bound of $\Omega(n^{1/3})$ for this task from~\cite{zhandry:setEquality}.
Indeed, let $p$ be a probability distribution and let $y$ be a fixed string encoding $p$ as in model (i).
Let $\sigma\colon [n]\to[n]$ be a function, and consider the input string $x$ given by $x_i = y_{\sigma(i)}$, which can be simulated given oracle access to $\sigma$ (as the string $y$ is fixed).
If $\sigma$ is a random permutation, then $x$ is a uniformly random input string from model (i).
If $\sigma$ is a random function, then $x$ is distributed as in model (ii).
\pfend

\section{Distinguishing Two Probability Distributions}

Recall the definition of Hellinger distance between two probability distributions $p$ and $q$ on the same space $A$:
\[
\dH(p,q) = 
 \sqrt{\frac12 \sum_{a\in A} \sA[\sqrt{p_a} - \sqrt{q_a}]^2}.
\]
Up to a constant factor, it equals $\|\mu_p - \mu_q\|$ and $1-\ip<\mu_p, \mu_q>$,
where $\mu_p$ and $\mu_q$ are as in~\rf{eqn:OracleII}.

In this section, we prove the following result:

\begin{thm}
\label{thm:main}
For any two probability distributions $p$ and $q$ on the same space $A$, and any model of accessing them from \rf{sec:models}, the quantum query complexity of distinguishing $p$ and $q$ is
\[
\Theta\sC[\frac{1}{\dH(p,q)}].
\]
\end{thm}

Note that this is quadratically better than complexity of the best classical algorithm \emph{for every choice of $p$ and $q$}.
Note also that for this problem model (iii) is equal in strength to the remaining models.

The proof of \rf{thm:main} involves proving lower and upper bounds in all four models, but, luckily, we can use relations from~\rf{prp:relations}.
The outline of the proof is as follows.
We prove upper bound in model (iv), which implies upper bounds in all other models as model (iv) is the most general of them.  
As for the lower bounds, we prove it for model (ii), which implies lower bounds in models (i) and (iv).  For model (iii), we prove the lower bound independently.
As a bonus, we prove an upper bound in model (iii) as a warm-up for the upper bound in model (iv).

In most of the proofs, we will use $\alpha$ for the angle between the vectors $\mu_p$ and $\mu_q$.  Note that
\[
\alpha = \Theta(\|\mu_p-\mu_q\|) = \Theta(\dH(p,q)).
\]

\subsection{Analysis in Model (iii)}
In this section, we analyse the problem in model (iii).

\begin{clm}
Quantum query complexity of distinguishing probability distributions $p$ and $q$ in model (iii) is $\Theta\sA[1/\dH(p,q)]$.
\end{clm}

\pfstart
Let us start with the upper bound.
Let $O$ be the input oracle, and let $U$ be a unitary that maps $\ket|\mu_p>$ into $\ket|0>$ and $\ket|\mu_q>$ into $\cos\alpha \ket|0> + \sin\alpha \ket|1>$.
Now use quantum amplitude amplification on the unitary $UO$ amplifying for the value $\ket|1>$.  
The algorithm can be also made exact using exact quantum amplitude amplification.

Now let us prove the lower bound.
Let $O_p$ be the input oracle exchanging $\ket|0>$ and $\ket|\mu_p>$ and leaving the vectors orthogonal to them intact.  Similarly, let $O_q$ exchange $\ket|0>$ and $\ket|\mu_q>$.  
Simple linear algebra shows $\|O_p - O_q\| = O(\alpha)$.  
(One way to see this is by using~\rf{eqn:RtoL2} and observing that $\|L_{\mu_p}-L_{\mu_q}\| = \|\mu_p - \mu_q\|$.)
Let $\EuScript{A}^O$ be a query algorithm making $t$ queries to $O$ and distinguishing $O_p$ from $O_q$.  Then,
\[
\norm| \EuScript{A}^{O_p} - \EuScript{A}^{O_q} | \le
t \|O_p - O_q\| = O(t\alpha).
\]
As this must be $\Omega(1)$, we get that $t = \Omega(1/\alpha)$.
\pfend

\subsection{Upper Bound in Model (iv)}
\label{sec:UpperI}

The aim of this section is to prove the following claim.

\begin{clm}
Quantum query complexity of distinguishing probability distributions $p$ and $q$ in model (iv) is $O\sA[1/\dH(p,q)]$.
\end{clm}

We prove this claim by constructing a feasible solution to~\rf{eqn:gamma2StatePreparing}.  In \rf{app:implementation}, we explain how to implement this algorithm time-efficiently.
In \rf{app:comparison}, we give a comparison to an algorithm using more typical techniques.
\medskip

Let $\psi$ and $\phi$ be some vectors encoding $p$ and $q$, respectively, as in model (iv).
That is,
\[
\psi = \bigboxplus_a \sqrt{p_a}\psi_a,
\qqand
\phi = \bigboxplus_a \sqrt{q_a}\phi_a,
\]
where $\psi_a$ and $\phi_a$ are some normalised vectors.
Our goal is to come up with a feasible solution to \rf{eqn:gamma2StatePreparing} with $\psi_x$ and $\psi_y$ replaced by $\psi$ and $\phi$.

We first analyse a pair of vectors $\sqrt{p_a}\psi_a$ and $\sqrt{q_a}\phi_a$ for a fixed $a$.
We would like to get a construction in the spirit of \rf{eqn:gamma2StatePreparing} that ``erases'' directions $\psi_a$ and $\phi_a$, and only depends on the norms $\sqrt{p_a}$ and $\sqrt{q_a}$.
One way is to use the following identity:
\begin{equation}
\label{eqn:trick}
\ipB<\sqrt{p_a}\psi_a, \sqrt{p_a}\psi_a-\sqrt{q_a}\phi_a> + \ipB<\sqrt{p_a}\psi_a-\sqrt{q_a}\phi_a, \sqrt{q_a}\phi_a> = p_a - q_a.
\end{equation}

We combine this identity over all $a$, add weights $c_a$, and re-normalise:
\begin{align*}
&\ipC< \frac{\bigboxplus_a c_a \sqrt{p_a} \psi_a}{\sqrt[4]{\sum_a c_a^2 p_a}}  ,\;
 (\psi-\phi)\cdot\sqrt[4]{\sum c_a^2 p_a} >
\\&\qquad+
\ipC< (\psi-\phi)\cdot \sqrt[4]{\sum c_a^2 q_a},\; \frac{\bigboxplus_a c_a\sqrt{q_a} \phi_a}{\sqrt[4]{\sum_a c_a^2 q_a}}>= \sum_a c_a (p_a - q_a),
\end{align*}
which gives
\[
\gamma_2 \sB[ \sum_a c_a (p_a - q_a) \midB L_\psi-L_\phi ]_{\psi,\phi} 
\le  
\sqrt{\sum_a c_a^2 p_a} + \sqrt{\sum_a c_a^2 q_a}.
\]
Dividing by $\sum_a c_a (p_a - q_a)$, we get that complexity of distinguishing $p$ from $q$ is at most
\begin{equation}
\label{eqn:complexityI}
O\s[\frac{\sqrt{\sum_a c_a^2 p_a} + \sqrt{\sum_a c_a^2 q_a}}{\sum_a c_a (p_a - q_a)}].
\end{equation}
Using triangle inequality
\[
\sqrt{\sum_a c_a^2 (\sqrt{p_a}+\sqrt{q_a})^2}
\le \sqrt{\sum_a c_a^2 p_a} + \sqrt{\sum_a c_a^2 q_a}
\le 2\sqrt{\sum_a c_a^2 (\sqrt{p_a}+\sqrt{q_a})^2},
\]
so~\rf{eqn:complexityI} is equivalent to
\[
O\s[ \frac{ \sqrt{\sum_a c_a^2 (\sqrt{p_a}+\sqrt{q_a})^2}}{\sum_a c_a (p_a - q_a)} ].
\]
Now it is easy to see that it is minimised to
\[
O\s[\frac1{\sqrt{\sum_a(\sqrt{p_a} - \sqrt{q_a})^2}}] = O\sC[\frac1{\dH(p,q)}]
\]
when $c_a = (\sqrt{p_a} - \sqrt{q_a})/(\sqrt{p_a} + \sqrt{q_a})$.

\subsection{Lower Bound in Model (ii)}
We use the following version of the adversary lower bound from~\cite{belovs:merkle}.

\begin{thm}
\label{thm:adv}
Assume $\EuScript{A}$ is a quantum algorithm that makes $T$ queries to the input string $x=(x_1,\dots,x_n)\in D$, with $D=A^n$, and then either accepts or rejects.
Let $P$ and $Q$ be two probability distributions on $D$, and $p_x$ and $q_y$ denote probabilities of $x$ and $y$ in $P$ and $Q$, respectively.
Let $s_P$ and $s_Q$ be acceptance probability of $\EuScript{A}$ when $x$ is sampled from $P$ and $Q$, respectively.  Then,
\begin{equation}
\label{eqn:adv}
T = \Omega\s[ \min_{j\in[n]} \frac{\delta_P^*\Gamma\delta_Q\* -  \tau(s_P,s_Q) \|\Gamma\|} {\|\Gamma\circ\Delta_j \|} ],
\end{equation}
for any $D\times D$ matrix $\Gamma$ with real entries. 
Here, $\delta_P\elem[x] = \sqrt{p_x}$ and $\delta_Q\elem[y]=\sqrt{q_y}$ are unit vectors in $\bR^D$;
for $j\in[n]$, the $D\times D$ matrix  $\Delta_j$ is defined by $\Delta_j\elem[x,y] = 1_{x_j\ne y_j}$; and
\begin{equation}
\label{eqn:dpq}
\tau(s_P,s_Q) = \sqrt{\strut s_P s_Q} + \sqrt{\strut (1-s_P)(1-s_Q)} \le 1 - \frac{|s_P-s_Q|^2}{8}.
\end{equation}
\end{thm}

In our case, $\delta_P = \mu_p^{\otimes n}$ and $\delta_{Q} = \mu_q^{\otimes n}$.
We construct $\Gamma$ as a tensor power $G^{\otimes n}$, where $G$ is an $A\times A$ matrix satisfying
\[
G \mu_q = \mu_p,
\qquad
\|G\|=1,
\qqand
\text{$\|G\circ\Delta\|$ is as small as possible,}
\]
where $\Delta$ is the $A\times A$ matrix given by $A\elem[a,b] = 1_{a\ne b}$.
Then,
\[
\delta_P^*\Gamma\delta_Q = \|\Gamma\| = 1,
\qqand
\|\Gamma\circ\Delta_j \| = \|G\circ\Delta\|,
\]
and \rf{thm:adv} gives the lower bound of $\Omega\sA[1/\|G\circ\Delta\|]$.

We construct $G$ as follows.  
Recall that $\alpha$ is the angle between $\mu_q$ and $\mu_p$.  
Then, $G$ is rotation by the angle $\alpha$ in the plane spanned by $\mu_q$ and $\mu_p$ and homothety with coefficient $\cos\alpha$ on its orthogonal complement.
That is, in an orthonormal basis where the first two vectors span the plane of $\mu_q$ and $\mu_p$, we have
\[
G =
\begin{pmatrix}
\cos\alpha & -\sin\alpha&0&\cdots&0\\
\sin\alpha & \cos\alpha&0&\cdots&0\\
0&0&\cos \alpha &\cdots &0\\
\vdots&\vdots&\vdots&\ddots&\vdots\\
0&0&0&\cdots&\cos\alpha
\end{pmatrix}.
\]
Clearly, $G\mu_q=\mu_p$ and $\|G\|=1$.
Let $G' = G-\cos\alpha\; I$.  We have
\[
\|G\circ\Delta\| = \|G'\circ\Delta\| \le 2 \|G'\| = 2\sin\alpha = O\sA[\dH(p,q)].
\]
For the inequality we used that $\gamma_2(\Delta)\le 2$, see~\cite[Theorem 3.4]{lee:stateConversion}.
This gives the required lower bound.

\section{Summary and Future Work}
In this paper we considered quantum algorithms dealing with classical probability distributions.
We identified four different models, and proved various relations between them.
We conjecture that models (i), (ii) and (iv) are equivalent.

Also, we considered the problem of distinguishing two probability distributions and obtained precise characterisation of its quantum query complexity in all four models in terms of Hellinger distance between the probability distributions.
The complexity turned out to be exactly quadratically smaller than the classical complexity of this problem for all pairs of distributions.

We showed that the corresponding algorithm can be implemented efficiently given that the probability distributions $p$ and $q$ can be handled efficiently.  We also compared our algorithm with a more standard approach using rejection sampling and amplitude estimation.

This raises a number of interesting open problems.
The first one is to prove or disprove the conjecture that models (i) and (iv) are equivalent.
Another interesting problem is to come up with a nice $\gamma_2$-characterisation of probability distribution oracles like \rf{thm:gamma2StatePreparing} characterises state-generating oracles.
Unfortunately, we do not have any hypothesis of how this characterisation might look like.
Finally, we would be interested in further quantum algorithms based on techniques of \rf{sec:UpperI}.

\subsection*{Acknowledgements}

Most of all I would like to thank Anr\'as Gily\'en for the suggestion to work on this problem.  I am also grateful to Fr\'ed\'eric Magniez, Shalev Ben-David, and Anurag Anshu for useful discussions.
Part of this research was performed while at the Institute for Quantum Computing in Waterloo, Canada.  I would like to thank Ashwin Nayak for hospitality.

This research is partly supported by the ERDF grant number 1.1.1.2/VIAA/1/16/113.

{
\small
\bibliographystyle{habbrvM}
\bibliography{belov}
}

\appendix

\section{Proof of \rf{thm:gamma2StatePreparing}}
\label{app:proofgamma2}

In this section, we freely use notions and results from~\cite{belovs:variations}, in particular Proposition 6 of that paper stating various properties of the relative $\gamma_2$-norm.

\begin{fact}
\label{fact:conjugate}
If $O_x$ are unitaries, then $\gamma_2 \sA[O_x^*-O_y^* \mid O_x - O_y] = 1$.
\end{fact}

\pfstart
The upper bound follows from $O_x^*(O_x - O_y) (-O_y^*) = -O_y^* + O_x^*$.  
The lower bound follows from the entry-wise lower bound property.
\pfend

\rf{thm:gamma2StatePreparing} follows from the following proposition.

\begin{prp}
\label{prp:stateEquivalence}
Let $O_x$ be a collection of unitaries in $\bC^{m+1}$, each preparing a state $\psi_x\in\bC^m$.  We have
\begin{equation}
\label{eqn:RtoL1}
\gamma_2\sA[L_{\psi_x} - L_{\psi_y} \mid O_x - O_y]_{x,y} \le 1.
\end{equation}
And conversely, let $\Psi$ be a collection of vectors in $\bC^m$.  Then, for each $\psi\in\Psi$, it is possible to define a operator $R_\psi$ preparing the state $\psi$ so that
\begin{equation}
\label{eqn:RtoL2}
\gamma_2\sA[ R_\psi - R_\phi  \mid  L_\psi - L_\phi]_{\psi,\phi} \le 3.
\end{equation}
Moreover, $R_\psi$ can be taken as the reflection through the orthogonal complement of $e_0 - \psi$.
\end{prp}

Let us explain how \rf{thm:gamma2StatePreparing} follows from \rf{prp:stateEquivalence}.
Let $f\colon D\to\{0,1\}$ be the function corresponding to the decision problem.  That is, $f(x)=0$ for all $x\in D_0$ and $f(y)=1$ for all $y\in D_1$.
By the striking out property, we have that
\[
\gamma_2(1\mid L_{\psi_x} - L_{\psi_y})_{x\in D_0, y\in D_1} \le 
\gamma_2(1_{f(x)\ne f(y)}\mid L_{\psi_x} - L_{\psi_y})_{x, y\in D}.
\]
On the other hand $\gamma_2\begin{pmatrix}
0&1\\1&0
\end{pmatrix} = 1$, and the duplication and the Hadamard product properties imply
\[
\gamma_2(1_{f(x)\ne f(y)}\mid L_{\psi_x} - L_{\psi_y})_{x, y\in D} \le
\gamma_2(1\mid L_{\psi_x} - L_{\psi_y})_{x\in D_0, y\in D_1},
\]
hence, these two quantities are equal.

Let $O_x$ be the oracle generating the state $\psi_x$ (formally, we duplicate $\psi_x$ for all possible $O_x$ generating the state).  Then~\rf{eqn:RtoL1} and the composition property imply that
\[
\gamma_2(1_{f(x)\ne f(y)}\mid O_x - O_y)_{x, y\in D} \le \gamma_2(1_{f(x)\ne f(y)}\mid L_{\psi_x} - L_{\psi_y})_{x,y\in D},
\]
proving the upper bound.
To prove the lower bound, we use~\rf{eqn:RtoL2} and the composition property:
\[
\gamma_2(1_{f(x)\ne f(y)}\mid L_{\psi_x} - L_{\psi_y})_{x,y\in D} \le 3\;\gamma_2(1_{f(x)\ne f(y)}\mid R_x - R_y)_{x, y\in D}.
\]

\begin{proof}[Proof of \rf{prp:stateEquivalence}]
Let us start with~\rf{eqn:RtoL1}.  First, we have
\[
\psi_x - \psi_y = (O_x - O_y) e_0
\qqhence
\gamma_2\sA[\psi_x - \psi_y \mid O_x - O_y] \le 1.
\]
Also, using \rf{fact:conjugate}, we have
\[
\psi_x^* - \psi^*_y = e_0^* (O_x^* - O_y^*)
\qqhence
\gamma_2\sA[\psi_x^* - \psi_y^* \mid O_x - O_y] \le 1.
\]
Using the direct sum property:
\[
\gamma_2\sA[(\psi_x - \psi_y)\oplus (\psi_x^* - \psi_y^*) \mid O_x - O_y] \le 1,
\]
which is equivalent to~\rf{eqn:RtoL1}.

Now let us prove~\rf{eqn:RtoL2}.  
%First we have to define $R_\psi$.
Note that $L_\psi^2$ is the projector onto the 2-dimensional space spanned by $e_0$ and $\psi$, hence, we can write
$
R_\psi = L_\psi + I - L_\psi^2.
$
%This is the reflection about the orthogonal complement of $e_0 - \psi$.
We have
\[
R_\psi - R_\phi = L_\psi - L_\phi + L_\phi^2 - L_\psi^2 
= (L_\psi - L_\phi) - (L_\psi - L_\phi) L_\phi - L_\psi (L_\psi - L_\phi).
\]
This gives~\rf{eqn:RtoL2} using triangle inequality for the relative $\gamma_2$-norm.
\end{proof}

\section{Low-Level Details of Implementations}

In this section, we sketch how the $\gamma_2$-bound from \rf{sec:UpperI} can be implemented as a quantum algorithm.
We first describe how to turn the general bound of \rf{thm:gamma2StatePreparing} into an algorithm, after that, the special case of \rf{sec:UpperI} is straightforward.

The implementations follow a standard routine of implementing this kind of algorithms, and rely on the following two results.

\begin{lem}[Effective Spectral Gap Lemma~\cite{lee:stateConversion}] \label{lem:effective}
Let $\Pi_A$ and $\Pi_B$ be two orthogonal projectors in the same vector space, and $R_A = 2\Pi_A-I$ and $R_B = 2\Pi_B-I$ be the reflections about their images.
Assume $P_\delta$, where $\delta\ge0$, is the orthogonal projector on the span of the eigenvectors of $R_BR_A$ with eigenvalues $\ee^{\ii\theta}$ such that $|\theta|\le \delta$.  Then, for any vector $w$ in the kernel of $\Pi_A$, we have
\[ \|P_\Theta \Pi_B w \|\le \frac{\delta}{2}\|w\|. \]
\end{lem}

\begin{thm}[Phase Estimation~\cite{kitaev:phaseEstimation, cleve:phaseEstimation}]
\label{thm:estimation}
Assume a unitary $U$ is given as a black box.  There exists a quantum algorithm that, given an eigenvector $\psi$ of $U$ with eigenvalue $\ee^{\ii\phi}$, outputs a real number $w$ such that $|w-\phi|\le\delta$ with probability at least $9/10$.  Moreover, the algorithm uses $O(1/\delta)$ controlled applications of $U$ and $\frac{1}{\delta}\polylog(1/\delta)$ other elementary operations.
\end{thm}

\subsection{Decision Problems with State-Preparing Oracles}
\label{app:gamma2StatePreparing}

Let us describe how to convert a feasible solution the optimisation problem of~\rf{eqn:gamma2StatePreparing} into a quantum algorithm that distinguishes $D_0$ from $D_1$.  This is a standard implementation of adversary-like algorithms.

The space of the algorithm is $\reg A \oplus \reg{BCXW}$,
where $\reg X$ and $\reg W$ correspond to the vector spaces $\cX$ and $\cW$, $\reg A$ is a 1-dimensional space, and $\reg B$ and $\reg C$ are qubits.

Let $T$ be the objective value of~\rf{eqn:gamma2StatePreparing}, and $0<\eps<1$ be the error parameter.
For each $x\in D_0$ define a vector
\[
\mu_x = \ket A|0> + \frac{\eps}{\sqrt{T}} 
\skC[{
    \ket B|0>\sB[ \ket C|0>\ketA XW|(O_x^*\otimes I)v_x> + \ket C|1> \ket XW|v_x> ] +
    \ket B|1>\sB[ \ket C|0>\ket X|0> + \ket C|1>\ket X|\psi_x> ]\ket W|u_x>
}].
\]
And let $\Lambda$ be the projector onto the span of all $\mu_x$.
For $z\in D$, let $\Pi_z$ be the projection onto the \emph{orthogonal complement} of
\[
I_{\reg B}\otimes \spn\limits_{\varphi\in \bC^{\reg X}}\, \sA[\ket C|0>\ket X|\varphi> - \ket C|1>\ket X|O_z\varphi>] \otimes I_{\reg W}.
\]
The reflection about $\Pi_z$ can be implemented in two queries to $O_z$.

The algorithm performs phase estimation subroutine on the operator $U = (2\Lambda-I)(2\Pi_z-I)$ and initial state $\ket A|0>$ with precision $\approx \eps^2/T$.
The algorithm accepts if the detected phase is 0.

In the positive case $z=x\in D_0$, there exists an eigenvalue-1 eigenvector of $U$ with large overlap with $\ket A|0>$, namely $\mu_x$.  So the phase estimation will report phase 0 with high probability.

In the negative case $z=y\in D_1$, we use the Effective Spectral Gap Lemma with witness
\[
w_y = \ket A|0> - \frac{\sqrt{T}}{\eps} 
\skC[{
    \ket B|0>\sB[ \ket C|0>\ket X|0> - \ket C|1>\ket X|\psi_y> ]\ket W|u_y> +
    \ket B|1>\sB[ -\ket C|0>\ketA XW|(O_y^*\otimes I)v_y> + \ket C|1> \ket XW|v_y> ] 
}].
\]
We have $\Pi_y w_y = \ket A|0>$, and, for all $x\in D_0$, $\ip <\mu_x, w_y>=0$.
Also,
\[
\|w_y\| = O\s[\sqrt{ 1 + T^2/\eps^2}] = O(T),
\]
since $T\ge 1$ and $\eps = \Theta(1)$.
By the Effective Spectral Gap Lemma, if we perform phase estimation with precision $\delta = \Theta(1/T)$, we reject with high probability.  This has complexity $O(1/\delta) = O(T)$.

\subsection{Implementation of the Algorithm}
\label{app:implementation}
Let us briefly describe how the feasible solution from \rf{sec:UpperI} can be implemented using techniques of \rf{app:gamma2StatePreparing}.
The main issue is to implement reflection about $\Lambda$.
Let the oracle register $\reg X$ be of the form $\reg D \oplus \reg{EF}$, where $\reg D$ is one-dimensional storing $\ket|0>$, $\reg E$ stores the index $a$, and $\reg F$ stores the vectors $\psi_a$.
The register $\reg W$ is one-dimensional in this case, so we will ignore it.

It is easy to see that $\Lambda$ is the span of the vectors of the form
\[
\ket A|0> + \ket BC |\nu_0>\ket D|0> + \sum_a \ket BC|\nu_a> \ket E|a>\ket F|\psi_a>,
\]
where $\nu_a$ are specific 4-dimensional vectors depending on $p_a$ and $q_a$, and $\psi_a$ are arbitrary vectors of the norm $\sqrt{p_a}$.
This space breaks down into a direct sum of orthogonal subspaces:
\[
\ket A|0> + \ket BC |\nu_0>\ket D|0>
\qqand
\ket BC|\nu_a>\ket E|a> \otimes I_{\reg F}.
\]
Hence, it is easy to perform reflection about $\Lambda$, given that all the vectors $\nu_a$ can be efficiently generated.

\section{Comparison to Standard Methods}
\label{app:comparison}
It is interesting to compare the bound as derived in \rf{sec:UpperI} to a more standard algorithm involving quantum rejection sampling as in~\cite{hamoudi:chebyshev}.

Assume we perform the following transformation involving the oracle in~\rf{eqn:OracleI}:
\begin{equation}
\label{eqn:comp1}
\ket|0> \longmapsto \sum_a\sqrt{p_a}\ket|a>\ket|\psi_a>\sB[\sqrt{1-c_a}\ket|0> + \sqrt{c_a}\ket|1>]
\end{equation}
or
\begin{equation}
\label{eqn:comp2}
\ket|0> \longmapsto \sum_a\sqrt{q_a}\ket|a>\ket|\phi_a>\sB[\sqrt{1-c_a}\ket|0> + \sqrt{c_a}\ket|1>],
\end{equation}
where $0\le c_a\le 1$ are some real numbers.
Then, we perform quantum amplitude estimation on the value $\ket|1>$ of the last register.
Since it is suboptimal to estimate simultaneously for $a$ such that $p_a<q_a$ and $a$ such that $p_a>q_a$, we may only consider those $a$ for which $p_a\ge q_a$.

Let $S_p$ and $S_q$ be probabilities of measuring 1 in the states from~\rf{eqn:comp1} and~\rf{eqn:comp2}, respectively.
Using standard bounds from quantum amplitude estimation~\cite{brassard:amplification}, we get that the required number of queries is
\begin{equation}
\label{eqn:alternative}
O\sC[\frac{\sqrt{S_p}}{S_p-S_q}] =
O\sC[\frac{\sqrt{\sum_a c_a p_a}}{ \sum_a c_a(p_a-q_a) }].
\end{equation}
Compared to~\rf{eqn:complexityI}, the difference is that there is no square at $c_a$ in the numerator, but there is restriction of $c_a\le 1$.  Since $c_a^2\le c_a$ for $c_a\le 1$, the bound of~\rf{eqn:alternative} can only be worse than the bound of~\rf{eqn:complexityI}.

Having no restriction on $c_a$ makes analysis simpler, and it is also possible to exhibit an example where the separation between~\rf{eqn:alternative} and~\rf{eqn:complexityI} is super-constant.
Let $t$ be a positive integer, $n=1+4+\cdots+4^{t-1}$, and $\alpha$ be a parameter we will specify later.
We will define probability distribution on $a\in[2n]$.
Let $p_a = \alpha$ for all $a\in[n]$.
As for $q_a$, they will be divided into $t$ consecutive groups of lengths $1,4,\dots,4^{t-1}$, respectively, such that all $q_a$ in the $i$th group equal $(1-2^{1-i})\alpha$.
Finally, for $i\in[n]$, let $p_{a+n} = q_a$ and $q_{a+n}=p_n$.
The scaling parameter $\alpha$ is chosen so that both $p_a$ and $q_a$ form a probability distribution.  Hence, $\alpha\approx 1/2n$.
\medskip

Let us first find the optimal value of
\[
\frac{\sqrt{\sum_a c_a^2 p_a}}{ \sum_a c_a(p_a-q_a) }
\]
for this distribution.  It is equal to
\[
\frac{\sqrt{\sum_a c_a^2}}{\sqrt{\alpha} \sum_a c_a w_a}
=\frac{\|u\|}{\sqrt\alpha \ip<u, w>},
\]
where $w=(w_a)$ is the vector
\[
\sB[1,\frac12,\frac12,\frac12,\frac12,
\frac14,\frac14,\frac14,\frac14,
\frac14,\frac14,\frac14,\frac14,
\frac14,\frac14,\frac14,\frac14,
\frac14,\frac14,\frac14,\frac14,
\dots ],
\]
and $u$ is the vector formed by $c_a$.
The optimal value is
\[
\frac1{\sqrt{\alpha}\|w\|} = O\sC[\sqrt{\frac{n}{\log n}}],
\]
which is achieved when $u=w$.
\medskip

Now let us consider the optimal value of
\[
\frac{\sqrt{\sum_a c_a}}{\sqrt{\alpha} \sum_a c_a w_a}
\]
subject to $0\le c_a\le 1$.
It is easy to see that in an optimal solution there exists $A$ such that $c_a=1$ for $a<A$ and $c_a=0$ for $a>A$.  Hence, up to a constant, the optimal value is equal to the minimum of 
\[
\frac{\|u\|}{\sqrt\alpha \ip<u, w>},
\]
where $u$ ranges over all vectors of the form $(1,1,\dots,1,0,0,\dots,0)$.

The vector $w$ breaks down into $t$ groups composed of equal entries.
The sum of the elements in each of the $t$ groups of $w$ is twice the sum of the elements of the preceding group, hence, the inner product $\ip<u,w>$ is at most 4 times the contribution of the last group fully covered by the ones in $u$.  
Thus, we have
\[
\frac{\|u\|}{\sqrt\alpha \ip<u, w>} \ge \frac{\sqrt{4^i}}{4\sqrt\alpha\cdot 4^i\cdot 1/2^i} = \Omega\sB[\frac1{\sqrt\alpha}] = \Omega(\sqrt n).
\]

\end{document}